# Spin Swapping for an Exchange Magnon Spin Current

Shuting Cui[1], Peng Yan[2], Wei Luo[1], Xiaofei Yang[1], Yue Zhang[1*]

1. School of Integrated Circuit, Huazhong University of Science and Technology, Wuhan, China

2. School of Physics and State Key Laboratory of Electronic Thin Films and Integrated Devices, University of Electronic Science and Technology of China, Chengdu 610054, China

*Corresponding author: yue-zhang@hust.edu.cn (Yue Zhang)

We propose the spin swapping effect for an exchange magnon spin current in a perpendicularly magnetized ferromagnetic medium with in plane anisotropy on the surface. The excitation of a magnon current flowing along an in-plane direction with an out-of-plane spin polarization leads to the generation of a secondary exchange spin current propagating along the out-of-plane direction, characterized by an in-plane spin polarization. The resulting exchange magnon spin current can induce an inverse spin Hall voltage of micro-volts. The exchange coupling at the interface between regions with different magnetic anisotropies plays a crucial role in generating the spin swapping effect. This is in contrast to the recently reported spin swapping for an exchange spin current in a canted antiferromagnet due to the Dzyaloshiskii-Moriya interaction.

The spin current is a key concept in spintronics, and the investigation of spin current paves a way for developing spintronic devices without serious Joule heat [1]. A spin current tensor comprises two vector components: spin polarization and the flow of (quasi-)particles carrying spins [2]. In 2009, M. B. Lifshitz and M. I. Dyakonov predicted the spin swapping effect in a ferromagnetic metal (FM)/nonmagnetic metal (NM) bilayer with spin-orbit coupling at the FM/NM interface, which involves the interchange of the directions for the two components of a spin current tensor [3]. This prediction has inspired extensive theoretical investigations in the spin swapping for transporting electrons in metallic systems over the past decade [3–7].

Besides electrons, magnons can also transport spins due to the conservation of angular momentum [1,2,8]. The magnon flow carrying a spin (magnetization in equilibrium) is referred to as an exchange magnon spin current, since it originates from the exchange coupling between neighboring magnetic moments [1]. Very recently, Lin et al. observed magnonic spin swapping for an exchange spin current in a canted antiferromagnetic (AFM) insulator and ascribed it to the Dzyaloshiskii-Moriya interaction (DMI) [9]. While the mechanism for the magnonic spin swapping remains controversial, there is a consensus that it predominantly appears in canted AFM insulators, such as $LaFeO_3$ and $LuFeO_3$ [9,10]. In other magnetic medium like a ferrimagnetic YIG, the spin swapping is negligible [10]. It remains an open issue whether the magnonic spin swapping is indeed a unique for a canted AFM medium.

Experimental verification of magnonic spin swapping was achieved by measuring the inverse spin Hall voltage ($U_{ISHE}$) which is proportional to the spin current density on the *surface* of a magnetic medium [11]. It is known that the magnetic structure near the surface can differ from the interior due to various factors, such as surface domain, dead magnetization layer, surficial anisotropy, and interfacial DMI [12–16]. This spatially inhomogeneous magnetic structure may enhance an exchange magnon spin current that is proportional to the spatial derivation of magnetization [1,2]. Therefore, in a medium exhibiting an inhomogeneous surficial magnetic structure, magnonic spin currents with interchangeable flow directions and spin polarizations are possible.

In this letter, we propose magnonic spin swapping in a perpendicularly magnetized *ferromagnetic* medium with surficial in-plane anisotropy (IPA). Our study demonstrates that because of the exchange coupling at the interface between two regions featuring different magnetic anisotropies, an exchange magnon spin current flowing along the out-of-plane direction with an in-plane spin polarization can be triggered by the inner spin wave flowing along an in-plane direction with an out-of-plane spin polarization, exhibiting typical spin swapping characteristics (Fig. 1). This surficial exchange magnon spin current can give rise to a micro-volt $U_{ISHE}$, comparable to that in a canted AFM medium [9]. However, the spin swapping in this work is different in that it originates from exchange coupling instead of DMI.

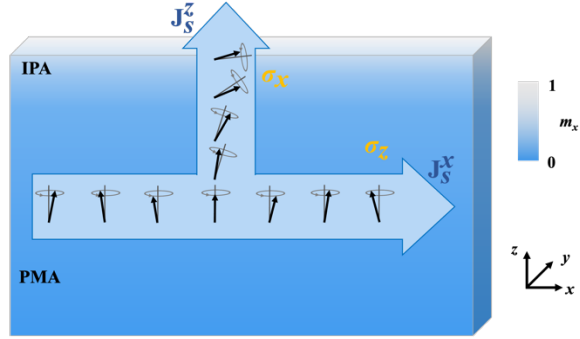

**Figure 1. Schematic of magnonic spin swapping for an exchange spin current in a perpendicularly magnetized medium with a surficial in-plane-anisotropy.**

The micromagnetic simulation was carried out by using a MUMAX software. To controllably modify surficial magnetic properties, we consider a ferromagnetic medium (saturation magnetization $M_S = 6.9 \times 10^5$ A/m) with inner perpendicular magnetic anisotropy (PMA) and a thin surficial layer with IPA. The easy axis of surficial IPA is along the $x$-axis direction with an anisotropy constant of $1 \times 10^4$ J/m$^3$, and that of PMA is $1 \times 10^6$ J/m$^3$. The dimension of the PMA layer is 400 nm × 200 nm × 100 nm, and the IPA layer ($t_{IP}$) has a thickness ranging from 0 and 12 nm. The cell dimension is 0.5 nm, which is smaller than the exchange length (~ 10 nm) as calculated by $l_{ex} = \sqrt{\dfrac{2A}{\mu_0 M_S}}$ and in close proximity to the lattice parameter. This ensures precise calculation of the spatial derivation of magnetization.

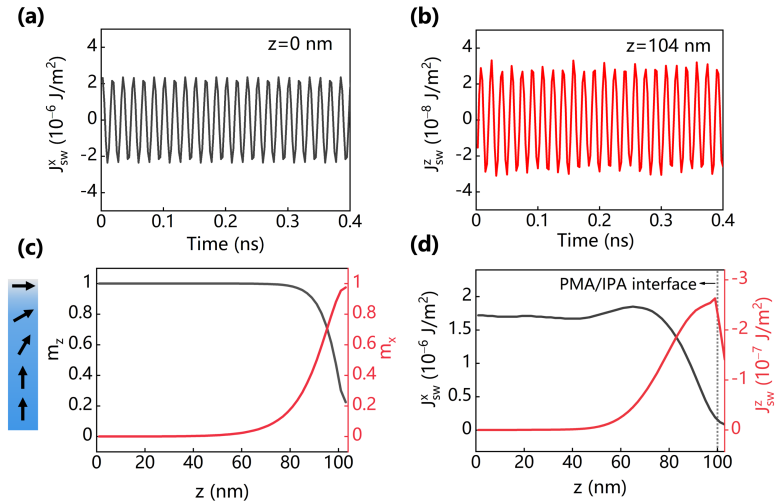

**Figure 2.** Temporal magnon spin current density (a) inner the PMA layer (z = 0 nm) and (b) on the surface (z = 104 nm). (c) Changing of $m_x$ and $m_z$ at the equilibrium state and (d) the exchange magnon spin current density $J_{sw}^x$ and $J_{sw}^z$ along the thickness direction.

A spin wave was excited by a localized alternating magnetic field $\vec{H}_{ac} = H_{max} \sin(2\pi f t)\vec{e}_x$ in the middle of the medium. Here $f$ and $H_{max}$ are the linear frequency and amplitude, respectively. Figures 2 (a) and (b) exhibit the magnetization oscillation inner the PMA medium (z = 0) and on the surface (z = 104 nm) for a 4-nm thick IPA layer. The magnetization direction transits from inner z-axis direction to surficial x-axis direction. Since the alternating field is along the x-axis direction, parallel to the IPA easy axis, we will demonstrate that the magnetic oscillation on the surface was not predominantly excited by the external field but triggered by the magnetization oscillation inner the PMA layer.

The density of an exchange magnon spin current flowing along the x(z) axis was quantified as $J_{sw}^{x(z)} = A[\vec{m} \times \frac{\partial \vec{m}}{\partial x(z)}]_{x(z)}$ [1]. Here the subscript sw is short for spin swapping. The spatial distribution of $m_x$ and $m_z$ in equilibrium and magnon spin current density were indicated in Figs. 2(c) and (d), respectively. Here we considered the spin current density at $t = 6$ ns when the magnetic oscillation is sufficiently stable. Inner the PMA layer (z = 0 nm), the magnon spin current composed by $m_z$ and $J_s^x$ [denoted by $(m_z, J_s^x)$] is dominant over $(m_x, J_s^z)$. On the surface and in the vicinity of the PMA/IPA interface (z = 100 ~ 104 nm), the magnon spin current $(m_x, J_s^z)$ assumes a prominent role. This interchange of the directions for two spin-current components indicates typical spin swapping characteristics.

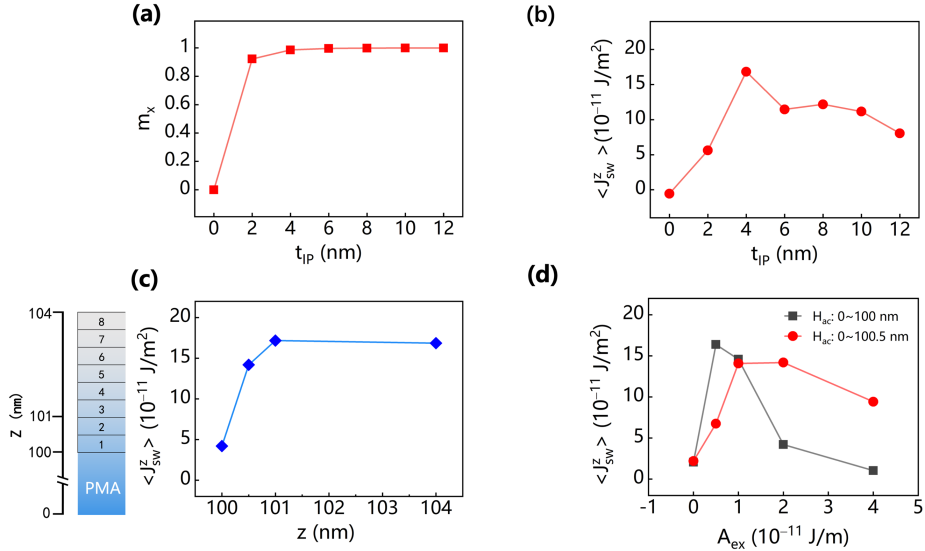

**Figure 3.** (a) $m_x$ in equilibrium as a function of the thickness of the IPA surficial layer ($t_{IP}$). (b) Average surficial magnon spin current ($<J_{sw}^z>$) as a function of the thickness of $t_{IP}$. (c) $<J_{sw}^z>$ under the exciting field acting on different ranges along the thickness direction from 0 to z. (d)

$<J_{sw}^z>$ **under different exchange constants at the PMA/IPA interface.**

In experiments, the magnonic spin swapping can be characterized by measuring a DC $U_{ISHE}$ that is proportional to the average surficial spin current density along the z-axis ($<J_{sw}^z>$). Here $<J_{sw}^z>$ was calculated by averaging $J_{sw}^z$ over a sufficiently long period to ensure a fixed $<J_{sw}^z>$. The equilibrium $m_x$ and the $<J_{sw}^z>$ on the surface were calculated at different $t_{IP}$ [Fig. 3(a) and (b)]. When the $t_{IP}$ is 4 nm or smaller, the $m_x$ significantly increases, which is accompanied with the obvious enhancement of $<J_{sw}^z>$, and it reaches $1.7 \times 10^{-10}$ J/m² at $t_{IP}$ = 4 nm. At a higher $t_{IP}$, the $m_x$ approaches 1, and $<J_{sw}^z>$ gradually decreases, but remains higher than $5 \times 10^{-11}$ J/m². Since $<J_{sw}^z> = A(<m_x \frac{\partial m_y}{\partial z}> - <m_y \frac{\partial m_x}{\partial z}>)$, the difference between $<m_x \frac{\partial m_y}{\partial z}>$ and $<m_y \frac{\partial m_x}{\partial z}>$ determines the magnitude of $<J_{sw}^z>$. When the IPA layer is sufficiently thick, the surficial magnetization stably aligns along the x-axis direction [Fig. 3(a)], and $m_x$ is significantly larger than $m_y$, giving rise to the domination of $<m_x \frac{\partial m_y}{\partial z}>$ over $<m_y \frac{\partial m_x}{\partial z}>$.

To confirm that the surficial spin current is generated by spin swapping instead of the excitation under external field, we restricted the spatial range of the alternating field from 0 to z [Fig. 3(c)]. When the field is acting on the PMA layer (z = 0 ~ 100 nm), the surficial $<J_{sw}^z>$ is approximately $4.2 \times 10^{-11}$ J/m². However, when the range of the field extended to cover a 1-nm IPA layer (z = 0 ~ 101 nm), the $<J_{sw}^z>$ has become almost the same as when the field covering the entire bilayer (z = 0 ~ 104 nm). This indicates that the surficial spin current is transmitted from the magnetic oscillation near the PMA/IPA interface.

To unravel the role of the exchange coupling at the PMA/IPA interface on the spin swapping, we calculated the surficial $<J_{sw}^z>$ for different exchange constants ($A_{ex}$) at the PMA/IPA interface [Fig. 3(d)]. Here the spatial range of excitation field is 0 ~ 100 nm and 0 ~ 100.5 nm. When this interfacial exchange coupling is absent ($A_{ex}$ = 0 J/m), the surficial $<J_{sw}^z>$ was significantly depressed. At a small $A_{ex}$, the surficial $<J_{sw}^z>$ obviously increases for both ranges of excitation field. This indicates that moderate interfacial exchange coupling allows the magnon flowing along an in-plane direction with an out-of-plane spin polarization to excite the secondary exchange magnon spin current flowing along the thickness direction with surficial in-plane spin polarization, thereby verifying the spin swapping behavior.

However, a larger $A_{ex}$ leads to suppress of the surficial $<J_{sw}^z>$ since strong interfacial exchange coupling drags the magnetic moments in the IPA surficial region away from the out-of-plane direction, hindering the spatial variation of magnetization.

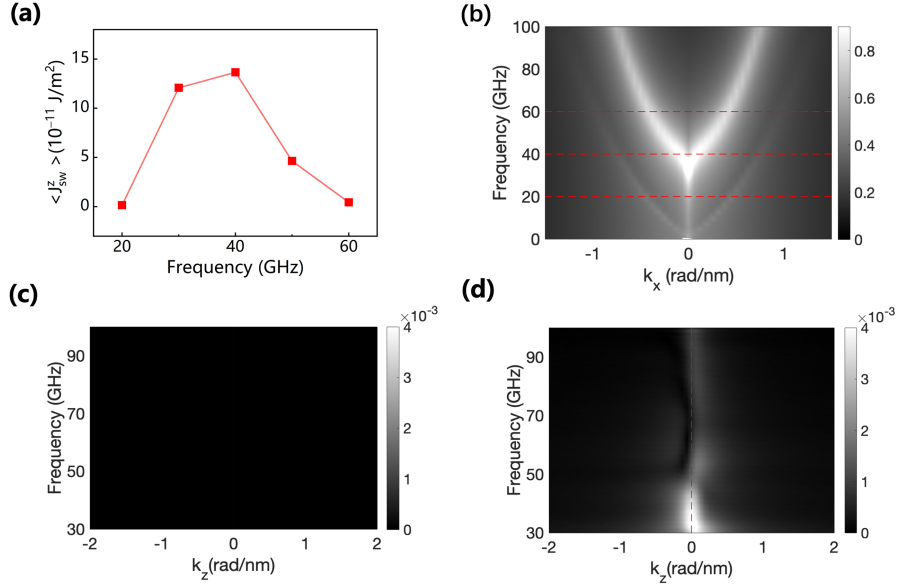

**Figure 4.** (a) Average $<J_{sw}^z>$ as a function of the excitation field frequency for a bilayer with $t_{IP}$ = 4 nm. (b) Dispersion relationship curve for a bilayer with $t_{IP}$ = 4 nm. (c) and (d) Dispersion relationship for the exchange magnon spin current flowing along the $z$ direction in the PMA layer for a single PMA medium and a bilayer with $t_{IP}$ = 4 nm, respectively.

$<J_{sw}^z>$ also exhibits non-monotonous changing with the frequency of excitation field, and maximum $<J_{sw}^z>$ appears at 40 GHz [Fig. 4(a)]. To obtain an understanding, we calculate the dispersion relationship curves of the bilayer for the exchange spin current flowing along the $x$-axis direction. It consists of two branches, a high-frequency PMA branch and a low-frequency IPA one [Fig. 4(b)]. It is noticed that the high-frequency branch exhibits a wide frequency range for a fixed wavelength, which is due to the exchange coupling at the PMA/IPA interface [17]. When the frequency is around the ferromagnetic resonance frequency ($f_{FMR}$) of the PMA layer, the $<J_{sw}^z>$ reaches its maximum value. This indicates that the excitation of a magnetostatic spin wave in the PMA layer can excite strong spin swapping. The $<J_{sw}^z>$ at a higher frequency is much smaller due to the difficulty for exciting an exchange wave. On the other hand, when the frequency is far below $f_{FMR}$, the $<J_{sw}^z>$ is also negligible even though this frequency is sufficiently high for the spin-wave transportation in the IPA layer. This further verifies that the $<J_{sw}^z>$ results from the spin swapping by the spin wave in the PMA layer instead of excitation under the external field.

Similarly, by applying the same field for exciting spin wave propagation along the *x*-axis direction, we derived the dispersion relationship of the exchange spin current flowing along the *z*-axis direction inner the PMA layer. In the single PMA medium, negligible exchange magnon spin current along the *z*-axis can be detected [Fig. 4(c)]. Conversely, in the bilayer with a 4-nm thick IPA layer, a weak exchange magnon spin current along the *z*-axis was generated [Fig. 4(d)]. This vertical spin wave exhibits a large wavelength, and exhibits a large strength in the frequency range between 30 and 50 GHz.

To experimentally characterize spin current, a heavy metal (HM) layer (like Pt) should be deposited above the magnetic layer, so that the spin current penetrating in the HM layer ($\vec{J}_s^z$) can be converted into a transversal electrical current ($\vec{J}_c$) through the ISHE effect and contributes to $U_{ISHE}$. On the other hand, the magnetization precession on the surface and at the PMA/IAP interface can also pump a spin current, but this spin pumping effect is still quite controversial [18–22]. Therefore, in the subsequent analysis, we focus on the estimation of the $U_{ISHE}$ resulting from spin swapping, and considered possible contribution from the spin pumping in the Supplementary Materials. We show that when the IPA layer reaches sufficient thickness, the spin current excited by spin pumping can be safely disregarded (S1 in the Supplementary Materials).

We estimated the $U_{ISHE}$ based on the equation [11]:

$$U_{ISHE} = \frac{2e}{\hbar}\frac{\lambda}{d}\theta_{SH}\rho w m_x <J_{sw}^z> \tanh(\frac{d}{2\lambda}).$$

Here *e* is the electron charge. We consider a Pt layer with a thickness *d* = 10 nm, a length *w* = 3 mm, and a width $\rho$ = 400 nm, located 50 nm away from the excitation source. Spin diffusion length $\lambda$ = 7 nm, and the spin Hall angle $\theta_{SH}$ = 0.08 [9]. The $U_{ISHE}$ indicates that for a surficial IPA layer with $t_{IP}$ = 2, 4, and 8 nm, the $U_{ISHE}$ is 1.04, 3.3, and 2.4 μV, respectively, close to the value reported by Lin et al [9].

In summary, we proposed magnonic spin swapping in a PMA/IPA bilayer. This spin swapping generates an exchange magnon spin current due to a moderate exchange coupling at the PMA/IPA interface. The spin swapping can be enhanced when the orientation of surficial magnetization gradually transits from the *z* to *x* axis and the frequency of the excitation field is within the range of a magnetostatic wave for the PMA phase. Additionally, this spin current can result in a $U_{ISHE}$ of several micro-volts. This work opens the door for realizing a magnonic spin swapping effect in a ferromagnetic medium with a mechanism that is distinct from that in canted AFM.

**Acknowledgment**
The authors acknowledge financial support from the National Key Research and Development Program of China (Grants No. 2022YFE0103300 and No. 2022YFA1402802) and the National Natural Science Foundation of China (Grants No. 51971098 and No. 12074057).